# Astro2020 Project White Paper

# PolyOculus: Low-cost Spectroscopy for the Community

**Thematic Areas:**  ☒ Planetary Systems   ☐ Star and Planet Formation
☒ Formation and Evolution of Compact Objects   ☒ Cosmology and Fundamental Physics
☒ Stars and Stellar Evolution  ☒ Resolved Stellar Populations and their Environments
☐ Galaxy Evolution   ☒ Multi-Messenger Astronomy and Astrophysics

**Principal Authors:**
Name: Stephen Eikenberry
Institution: University of Florida
Email: eiken@ufl.edu
Phone: 352-294-1833

**Co-authors:** (names and institutions): Misty Bentz (Georgia State University), Anthony Gonzalez (University of Florida), Joseph Harrington (University of Central Florida), Sarik Jeram (University of Florida), Nick Law (University of North Carolina – Chapel Hill), Tom Maccarone (Texas Tech University), Robert Quimby (San Diego State University), Amanda Townsend (University of Florida)

**Abstract**: As astronomy moves into the era of large-scale time-domain surveys, we are seeing a flood of new transient and variable sources which will reach biblical proportions with the advent of LSST. A key strategic challenge for astronomy in this era is the lack of suitable spectroscopic followup facilities. In response to this need, we have developed the PolyOculus approach for producing large-area-equivalent telescopes by using fiber optics to link modules of multiple semi-autonomous, small, inexpensive, commercial-off-the-shelf telescopes. Crucially, this scalable design has construction costs which are >10x lower than equivalent traditional large-area telescopes. In addition, PolyOculus is inherently highly automated and well-suited for remote operations. Development of this technology will enable the expansion of major research efforts in the LSST era to a host of smaller universities and colleges, including primarily-undergraduate institutions, for budgets consistent with their educational expenditures on similar facilities. We propose to develop and deploy a 1.6-m prototype demonstrator at the Mt. Laguna Observatory in California, followed by a full-scale 5-meter-class PolyOculus facility for linkage to existing and upcoming time-domain surveys.

## I.      Key Goals and Objectives:

As astronomy moves into the era of large-scale time-domain surveys, we are seeing a shift in both scientific focus and observing strategies for the coming decades. The Large Synoptic Survey Telescope (LSST) is the highest priority for US ground-based astronomy for the next decade. Meanwhile, other surveys have already begun to produce tremendous amounts of scientific data on rare and transient events (i.e. the Zwicky Transient Facility [1] and its precursor Palomar Transient Factor [2], Pan-STARRS [3], SkyMapper [4], CRTS [5], ATLAS [6], Evryscope [7], and many others). These facilities are already generating a flood of new transient and variable sources — a flood which will reach biblical proportions with the advent of LSST.

A key strategic challenge for astronomy in this era is the lack of suitable spectroscopic followup facilities. Followup spectroscopy will be scientifically crucial for identifying the basic nature of these sources (supernova? Flare star? tidal disruption event? new class of object?), their key physical properties (temperature, composition, radial velocity, outflows, etc.) and the evolution of those properties with time. However, traditional spectroscopic facilities are expensive to construct and to operate. Furthermore, even with the large rate of discoveries, the relative density of targets on the sky is quite low – typically <<1 per square degree per night (even for LSST). Thus, the multi-object spectroscopic facilities developed in recent decades provide little or no advantage for most time domain followup. This can be seen directly in Astro2020 Science White Papers, such as [8] and [9].

In response to this need, we have developed a method for producing large-area-equivalent telescopes by using fiber optics to link modules of multiple semi-autonomous, small, inexpensive, commercial-off-the-shelf (COTS) telescopes. Crucially, this scalable design has construction costs which are >10x lower than equivalent traditional large-area telescopes. This innovative "PolyOculus" array approach represents a transformational technology for spectroscopic followup in time-domain astronomy. In addition, PolyOculus is inherently highly automated and well-suited for remote operations. Development of this technology will enable the expansion of major research efforts in the LSST era to a host of smaller universities and colleges, including primarily-undergraduate institutions, for budgets consistent with their educational expenditures on similar facilities.

We propose to develop and deploy a 1.6-m prototype demonstrator at the Mt. Laguna Observatory in California, followed by a full-scale 5-meter-class PolyOculus facility for linkage to existing and upcoming time-domain surveys (including, but not limited to, Evryscope and LSST for a Southern Hemisphere deployment). This will provide world-class scientific results, while simultaneously demonstrating a low-cost, high-performance spectroscopic capability well-suited for replication and use by a large number of even small-/medium-sized institutions which might not otherwise have access to such facilities.

## II. Technical Overview

In this section, we describe the potential for PolyOculus to provide low-cost spectroscopic telescope facilities to address the needs noted above. We then use a strawman science case as an example for developing the top-level performance requirements for a PolyOculus system. We then move to top-level design requirements and a brief summary of the key characteristics of the detailed design we developed to meet those requirements.

### A. Low-cost spectroscopic collecting area in the 2020s

The strategic gap between the massive flood of imaging data in the LSST era and the relative scarcity and cost of relevant spectroscopic followup capabilities provides a key driver for this white paper: developing a low-cost, easily-replicated spectroscopic facility for public access and/or widespread private access even for small-/medium-sized institutions with limited budgets. Based on our calculations and simulations, and confirmed with our own laboratory tests and published results on fiber coupling efficiency from other groups, the net performance of the PolyOculus approach for a 1.6-meter-equivalent aperture is similar to that of a "standard" 1.6-meter telescope with an additional throughput loss of ~15% (sensitivity reduction of ~7-8%). Meanwhile, the total cost of such a PolyOculus system is under $600,000 (including facility/enclosures, but not counting the science instrument). For comparison, recent commercial quotes for a standard telescope from established vendors, including enclosure and guiding system range from $2,600,000 to $4,000,000 – see Figure 1. In other words, with a slight reduction in sensitivity, PolyOculus can reduce costs for telescope apertures to ~20% of the cost of standard telescopes – and for larger apertures this can reduce to ~10% due to an economy of scale. PolyOculus also has an important scalability advantage — an initial small array can be incrementally augmented with additional modules later on, without a "restart" of the telescope.

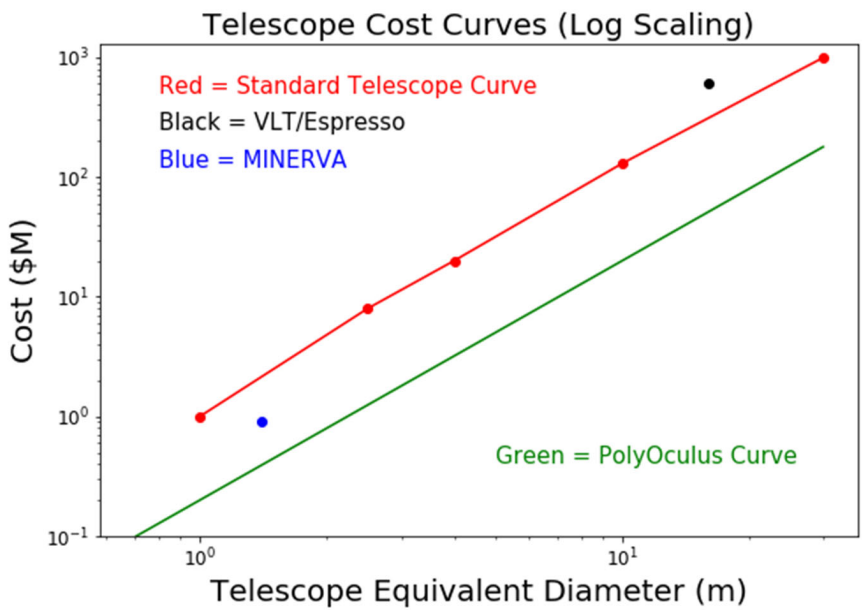

*Figure 1* – Comparison of telescope costs (with enclosures, but no instruments) for standard telescopes vs PolyOculus arrays. Prices are current-year dollars, drawn from various sources via private communication. Note that MINERVA [10] also provides imaging, but this is a "fair" comparison for spectroscopy-only.

### B. Key Performance Requirements

The performance requirements for any scientific experiment depend critically on the science use case. For the purposes of this white paper, we developed a "strawman" science case, extrapolated from the science descriptions in the Astro2020 white papers listed above, and augmented them with detailed science cases from our team members. We omit those details here (for space reasons), and simply note that they cover the gamut from explosive transients (supernovae, kilonovae, etc.) to X-ray binaries to active galactic nuclei to stellar populations to exoplanets. From those science cases above, we extracted the following Table of Top-level scientific performance requirements for this implementation of PolyOculus.

Table 2 – Top-level scientific performance requirements

| Parameter | Value | Science Drivers |
|---|---|---|
| Wavelength coverage (requirement) (1a) | 450-900 nm (req) | Blue end by Bowen blend; red end by red-shifted features from explosive transients |
| Wavelength coverage (goal) (1b) | 400-1000 nm (goal) | Blue end set by atmosphere/fiber transmission; red end set by Si band gap for sensitive low-cost detectors |
| Spectral res. (2a) | R~200 | Maximum sensitivity to explosive transients with broad lines |
| Spectral res. (2b) | R~500-1000 | Higher-res mode for bright sources: Stellar line IDs; large-amplitude radial velocity shifts (SS433, X-ray binary orbits); P-Cyg profiles for fast outflows; AGN reverberation |
| Sensitivity (requirement) (3a) | >22.7 mag, 10σ (1 hr) | Exceeds Evryscope sensitivity for single-night detection; good match to SNe in the local Universe |
| Sensitivity (goal) (3b) | >23 mag 10σ (1 hr) | Allows tracking of fading transients over multiple nights |
| Fiber field-of-view (4) | 1.2-1.6-arcsec | Optimal SNR for slit losses in typical seeing |
| Sky Coverage (5) (% of accessible sky) | >50%(requirement) 100% (goal) | Maximimize ability followup transients over full sky |
| Target multiplexing (full-sky) | 2 (req), 3+ (goal) | Maximizes observing efficiency for low target density; 2 required and 3+ goal for exoplanet atmospheres |
| Operation mode (6) | Remote; semi-autonomous | Required for low-cost operation; Also important for observing efficiency and rapid response |

We note that this approach is inherently very flexible, and other design options/outcomes are possible for other science drivers. For instance, we also have developed design studies for a 5-meter PolyOculus with a moderate resolution spectrograph, and another for a high-resolution "Cosmic Accelerometer" (see Astro 2020 white papers [11] and [12]). These could even share the same telescope facility, with a simple fiber switch allowing remote instrument switching in <60 seconds at the touch of a button. One possibility for an alternate instrument would be a low-cost high time-resolution photometer to address key science cases as described in the Astro2020 White Paper by Maccarone et al. [13].

### C. Observatory architecture/design

From the top-level science requirements in Table 2, we constructed a set of >60 performance requirements at the subsystem and component level (omitted here for space reasons), to guide the design. Our team has developed a complete design including the telescope array and the spectroscopic detector unit. We have used our previous test results, combined with a number of key trade studies on technologies and design approaches, to arrive at a design that meets all of the top-level requirements, many of the top-level science goals, while minimizing cost and mitigating technical risk. The end result is a robust, optimized design - we present a brief high-level synopsis in Table 3.

**Table 3 – Brief summary of key facility design features**

| Feature | Key Design Features | Driving Requirements |
|---|---|---|
| Unit telescope | Meade 14-inch LX-200 | Lowest cost per sq. meter (once including per-unit costs of AnG, ADC, control computer, etc.) |
| Telescope Array size | 28x PolyOculus 7-pack | Sensitivity needed for faint targets; Multiplexing for exoplanet transit atmospheric calibrations and simultaneous observations of multiple targets widely separated on the sky |
| Atmospheric Dispersion Corrector (ADC) | 7-arcsec throw (400-1000nm) 22-mm CA, 20-mm length | Maintain broadband throughput @2 airmass without increasing fiber aperture/sky-background; FOV for AnG guiding and sky coverage; max length to minimize mount interference (sky coverage issue), flexure, etc |
| Acquisition and Guide System (AnG) | 0.44-"/pix; 6'x6' FOV (Hartmann mode); g>16.0 mag in 60s | Pixel scale matches MLO seeing; FOV + sensitivity for sky coverage; focus tracking; lightweight/compact to minimize mount interference (sky coverage issue), flexure, etc |
| Fiber Feed | 1.2-arcsec FOV @ f/5; 10/150-micron broadband fiber | Wavelength coverage; match FOV to seeing to maximize sensitivity; match NA of fiber for etendue issues (see below) |
| Pointing/guiding control | full-sky pointing <4'; blind offset for targets <1"; guide <0.5" RMS | Target rough acquisition onto AnG FOV; blind offset from reference star in FOV onto fiber for faint targets (g>17 mag); guiding to maintain throughput/sensitivity |
| Fiber coupling | Pupil-packing @75% fill factor; switchable outputs, incl. spectral calibration lamps | Switchable outputs for multiple spectral resolutions + target multiplexing; packing efficiency for etendue conservation (cost/performance driver – see below); lamps for wavelength/flatfield calibration; power monitor for acquisition/throughput verification |
| Spectrograph | R ~500/1000 double spectrograph; rebin to R ~200; red/blue VPH gratings; bkgd-limited @>600nm | Broad wavelength coverage; high efficiency for sensitivity; higher spectral resolution for stellar science cases; low-resolution for extragalactic transients |
| Control Software | LBJEC/LBJDD Java agents + INDILIB | Remote operation; semi-autonomous function |

In the following subsections, we present a description of the design in more detail. We begin with the individual telescope units, then move to the 7-pack modules. We then present the full array and spectrograph unit. Finally, we conclude with a description of the key operational considerations.

1. PolyOculus basic telescope unit

We present the individual telescope unit for this approach in Figure 2. This unit is the basic "building block" of the PolyOculus array used for the facility we propose here. We selected the Meade 14-inch LX-200 ACF as the base telescope for PolyOculus (at this stage), due to its very low cost per square meter, combined with reasonable performance for optics, pointing, and tracking. The large field of view is overkill (we only use ~6-arcmin), and future arrays may use cheaper custom optics – but for the current economy of scale, we rely only on COTS systems. We also investigated other COTS options (such as PlaneWave), which seem to have better performance/reliability. However, their cost per unit area is 2x that of the Meade, which seem to perform "good enough" based on our tests of performance and reliability.

The focal plane package includes the fiber-optic feed, the Acquisition and Guide (AnG) system, an Atmospheric Dispersion Corrector (ADC), and a microfocuser. These are all either COTS systems, or low-cost designs based on COTS components which we have successfully prototyped and tested.

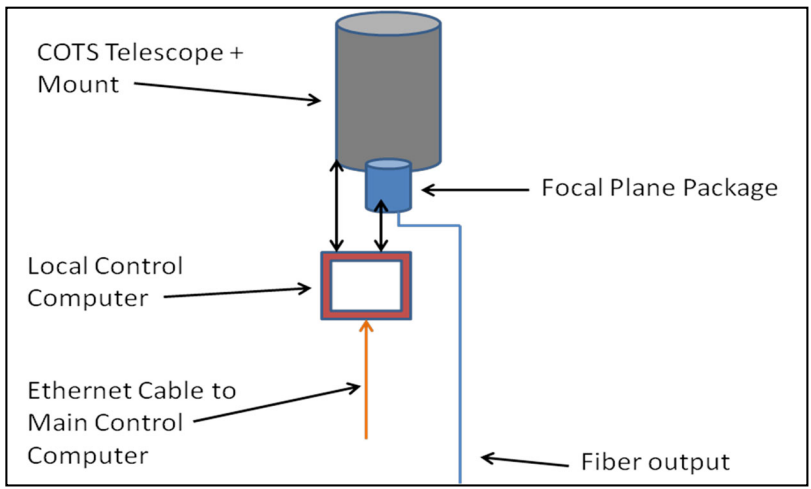

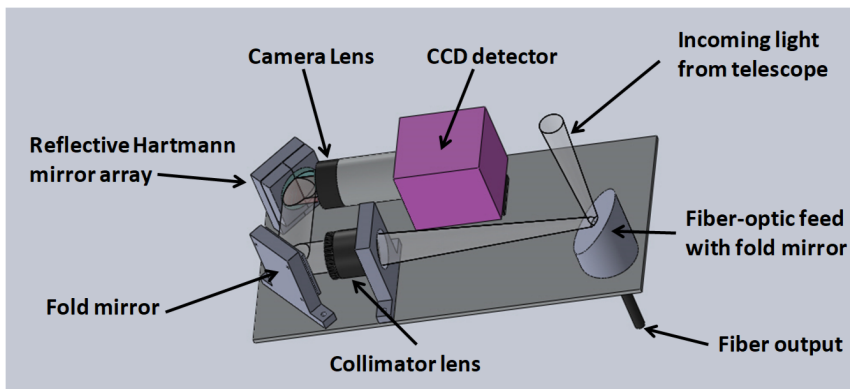

*Figure 2* – (Top) Basic individual telescope unit. The telescope and mount are commercial-off-the-shelf (COTS) technology. The focal plane package includes the ADC, fiber-optic feed, acquisition and guiding unit, and focus control (see text for further details). (Bottom) View of the fiber-optic feed and AnG system portion of the Focal Plane Package.

## 2. PolyOculus 7-Pack module

The basic module of the PolyOculus array is a "7-pack" — 7 individual telescope units with a fiber optic link (Figure 6). The 7 optical fiber feeds of the units are arranged in a "heptapack" fiber bundle. The heptapack arrangement is then placed at the optical input of a novel Switchable Multi-Fiber Coupler (SMFC) as shown in Figure 3. The SMFC takes the 7-pack fiber light and outputs it selectably to one of two fiber outputs or an Intensity Meter port.

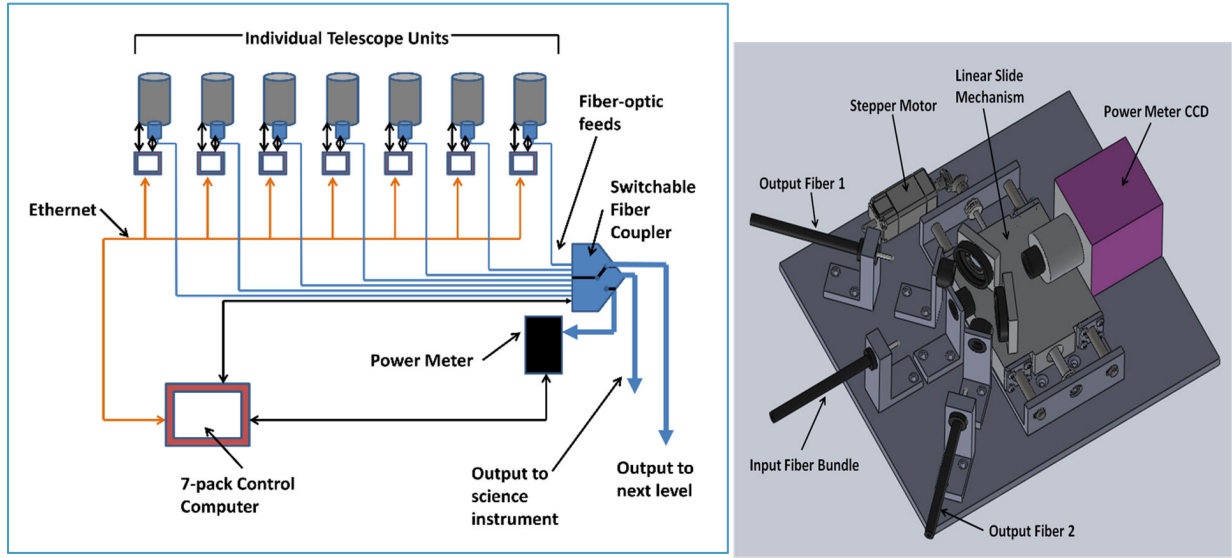

***Figure 3*** — *(Left) Conceptual layout of a basic "7-pack". The 7 individual telescope units are identical to the one shown in Figure 4, and operate autonomously for tracking and focus control. The 7 individual fibers transmit the light from the primary target on each telescope into a control enclosure housing the rest of the system optics and electronics. In the enclosures, the fibers are grouped together into a "heptapack" geometry. This geometry provides the maximum packing density for the fiber bundles.* (Right) *Switchable Multi-Fiber Coupler (SMFC).*

The approach we use for PolyOculus is a collimator/camera optical relay with a mechanically-selectable fold mirror, allowing us to direct the optical beam to either of two fiber outputs, or pass between the two fold mirrors to an output port which focuses a magnified image of the heptapack input onto the Intensity Meter – another COTS CCD. The Intensity Meter is used to autonomously check/verify the pointing of each individual telescope unit by making an image of the bundle output, showing the illumination of each fiber. The first of the two fiber outputs is connected to the scientific detector unit (described below), while the second fiber output directs the primary target light to the next level of the PolyOculus system. The SMFC also allows injection of light from calibration sources (arc and flat lamps) for calibration of the spectrograph.

The fill factor of the fiber packing at the SMFC input has an important impact on PolyOculus "downstream" performance. Any packing efficiency <100% requires that the entendue of the next optic in the system increase by 1/(fill factor). For spectrographs, this will increase the required pupil image size (and thus require larger, more expensive gratings and collimator/camera optics). For this PolyOculus implementation we will use "pupil plane packing", where the 7 input fibers are each coupled to a lenslet array creating a pupil plane with the pupil images closely packed together. The lenslet array is designed to maximize the fill factor of the pupil plane - the individual pupil image diameters match the lenslet pitch in the array, which in turn matches (or slightly exceeds) the clad diameter in the input fiber plane. While the fiber clad diameter thus influences the lens array, this approach retains flexibility via the focal length of the lenslets, which is straightforward to adjust to match the desired design. Furthermore, the packing efficiency of lenslets can be quite high – several reputable vendors report regularly achieving ~75% (close to the theoretical maximum of 78%).

   3. *PolyOculus full array*

The 7-pack PolyOculus module synthesizes an aperture equivalent to that of a telescope with 0.9-meter diameter — a substantial increase, and quite large compared to typical low-cost COTS telescopes. Furthermore, the PolyOculus approach is scalable to larger apertures. In this case, we require 28 of the 7-pack units combined into a single-output 5m-equivalent full array for sensitivity/multiplexing. This can be sub-divided into 4x 2.5m-equivalent arrays, to simultaneously observe 4 targets widely separated on the sky.

Following the approach above for the Level-1 SMFC above, the Level-2 SMFCs takes the 7 output fibers of the Level-1 SMFCs and couples them to lenslet arrays. These can then be fed directly into the spectrograph for science observations, or can be relayed to the Level-3 SMFC. At the Level 3 SMFC, the 4 outputs are relayed into a single output fiber, which can be fed into the spectrograph. Note that the other Level-3 SMFC could also feed an alternate science instrument with no additional complexity for the PolyOculus system. This allows "hot switching" between instruments on a timescale of <60 seconds, with no physical movement of instruments, etc. – a significant simplifying advantage for operations.

Each telescope in the PolyOculus array will be mounted on its own pier, with each 7-pack in a separate enclosure. The pier consists of a concrete COTS Hurricane pad, with attachments points for 3 "feet" via concrete anchors. The feet attach to the base of the steel pier support structure, also providing tip/tilt adjustment for leveling via bolt adjustments. All aspects of the pier system are designed for performance, robustness, and low manufacturing cost, with relative ease of assembly and transport. The enclosure is a custom-designed roll-off roof structure. The main structure consists of treated plywood with a 2x4-wooden frame. The footprint of the enclosure allows access to the telescopes for service/maintenance and operation without interference, and provides an unobstructed view of the sky to 2.0 airmasses. The roll-off roof rides on COTS garage door rails and is driven by a remotely-operated COTS electric sliding gate opener with a rack actuator. Access to service the telescopes is done through a hinged door at one side of the enclosure (and with the roof open). Each enclosure will also have an air conditioner and dehumidifier to prevent condensation on the exposed optics of the telescope.

The PolyOculus control software is based on our working prototype developed under the EAGER grant, built upon the Instrument Neutral Distributed Interface (INDI) library — a cross-platform software library designed to control a wide variety of astronomical instruments.

   4. *Science spectrograph design*

The spectrograph is the primary science detector unit for the array, and thus plays a key role in the scientific productivity of this facility. We have carried out a number of trade and design studies to optimize the scientific performance, while balancing against cost. A key feature of spectrographs for PolyOculus arrays is the availability at the spectrograph input focal plane of different fiber sizes and f/# beams coming from either the full array or the various sub-arrays. Another key issue for this spectrograph is the broad wavelength coverage requirement from the science cases above. Since the goal wavelength range of 400-1000nm exceeds a factor of 2, a single first-order grating is not practical, and our trade studies showed that the high-resolution goal was not easily compatible with prism dispersers. Thus, we turned to a standard double-spectrograph approach, with blue and red channels separated by a dichroic. The mechanical design is straightforward, as the spectrograph has no moving parts.

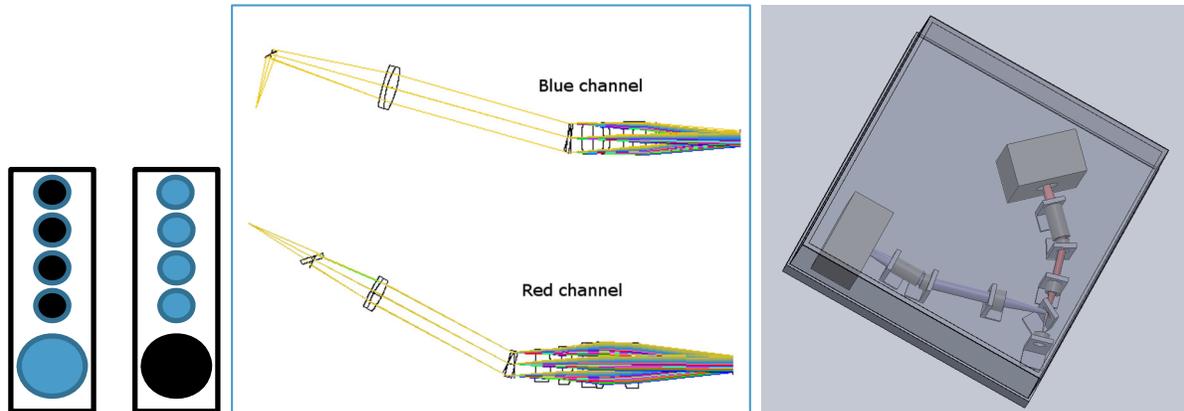

***Figure 4*** *– (Left) Sketch of the input focal plane for the spectrograph. The left box shows the full-array fiber illuminated, while the right box shows the 4x sub-array fibers illuminated. (Middle) Blue/red channel ZEMAX layouts. (Right) Preliminary mechanical design of the spectrograph on a breadboard.*

5. *Observatory operational model*

The operational approach for the observatory will be as a remotely-operated semi-autonomous facility. We will use an automated algorithm for target selection/scheduling, with daily human review/approval of the night's observing plan. The algorithm will give top priority to transients requiring rapid response, followed by high-priority monitoring observations, and ending with "filler" observations drawn from non-time-critical surveys. These will be loaded into the scheduling manager before the evening observing begins. The observing plan will also include offset stars for blind pointing acquisition on targets fainter than g~17 mag. The baseline will be for operations to be controlled from Florida and other remote sites. On-site staff will be available to assist in emergency situations. Routine maintenance and service will be carried out by UF staff on quarterly visits. We note that for PolyOculus, individual telescope failures do not stop operations – they simply reduce the effective collecting area. Our MTBF analysis indicates that quarterly visits will allow the observatory to operate at an average of >90% efficiency.

## III. Technology Drivers – PolyOculus Status and Development Needs

As noted in Section I, our primary proposal here is to develop and deploy a 1.6-m prototype demonstrator at the Mt. Laguna Observatory in California, followed by a full-scale 5-meter-class

PolyOculus facility. In the subsections below, we describe the key development plans and timescales for this work.

We have previously deployed an array of four telescopes outfitted with custom acquisition and guide units with fiber optic feeds for on-sky testing. We developed an all-sky pointing model for the telescopes that meets the requirement that autonomous pointing of the telescopes brings targets within the acquisition field-of-view for full-sky pointing. We also demonstrated that individual 14-inch telescope units can guide on stars with r>16 mag using the low-cost uncooled AnG CCD. We also developed and tested the required fiber linkage, again demonstrating low cost and reliable, quality performance.

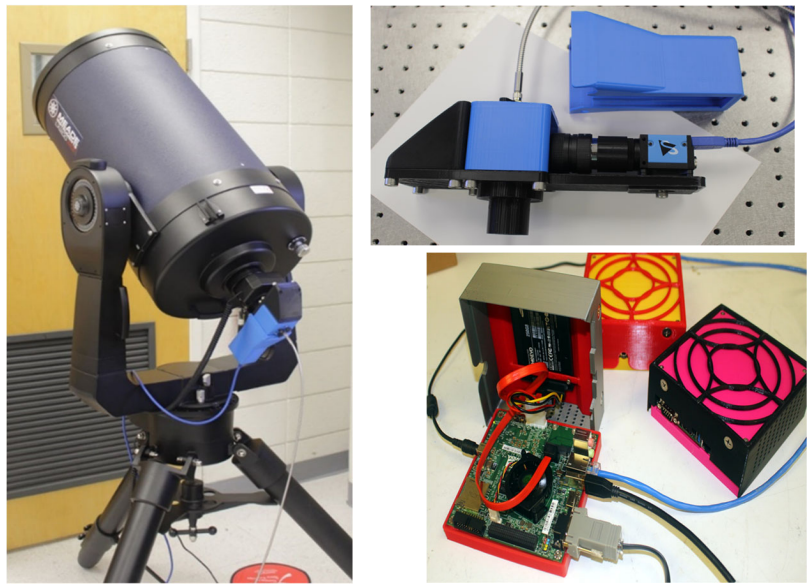

*Figure 5 – (Left) Meade 14-inch LX-200 ACF with prototype AnG unit and fiber feed. (Top Right) Prototype AnG unit with cover removed on CCD/lens. (Bottom Right) Jetson control computer*

Moving to a 5m-equivalent PolyOculus will take 5 years. During the first two years we will first demonstrate key on-sky performance parameters of a full PolyOculus array, including:

1. Fully-autonomous acquisition, guiding, and focusing while coupled to a science-grade spectrograph.

2. Multi-target spectroscopy for wide-separated science targets.

To achieve this demonstration, we will deploy a 1.6-m "Phase 0" demonstrator of the PolyOculus array. This array will be co-sited with Evryscope-North at the Mount Laguna Observatory in southern California. It will also include a small, low-cost double spectrograph as its scientific detector unit with spectral resolution modes from R~200 to R~1200. This demonstrator will provide low-latency spectroscopic identification of any transient detected by Evryscope in <1 hour, and follow more numerous fainter transients detected by other transient discovery facilities down to >20 mag. It will also provide critical spectroscopic follow-up for a broad range of Evryscope, TESS, and other investigations already underway, ranging from AGN to stellar physics to exoplanets. The 1.6m aperture consists of three 7-pack modules, allowing for the multi-target spectroscopy demonstration, while also providing acceptable overall sensitivity for the planned scientific observations. This effort during the first 24 months will effectively retire the risk for the full system related to the PolyOculus technology. The estimated cost of this development is ~$800K, including hardware (array + spectrograph), personnel, deployment, and a 6-month scientific demonstration phase. We note that this demonstrator is identical to the one proposed in [12], and could serve both purposes simultaneously.

## IV. Organization, Partnerships, and Current Status

Design and development of PolyOculus is led out of the University of Florida. Team members at Florida have longstanding expertise designing facility-class astronomical spectrographs as well as the PolyOculus prototypes thus far. The team also includes the leaders of the Evryscope facilities at CTIO and Mount Laguna Observatory. The science team includes expertise ranging from explosive transients to exoplanets to AGN to X-ray binaries to cosmology. This project is currently at the R&D level, but is ready to progress to deployment of the first 1.6m demonstrator system. The fiber-coupling technique has been demonstrated, as has the software control of the telescope array. No formal partnerships are in place at this time.

## V. Schedule & Cost Estimates

The basic outline of our proposed schedule is as follows. We anticipate a 2-year initial phase for the preliminary 1.6m demonstrator. This will be followed by a 3-year development and deployment period for the additional PolyOculus units to reach 5m Phase I array. This could either be at Mt. Laguna, as a scale-up of the demonstrator, or a separate facility in the Southern Hemisphere (allowing full sky coverage overlap with LSST). We then baseline a 3-year science operations period. This gives a total timeline for the proposed activities of 8 years.

For our cost estimation, we used the same costing analysis used at the University of Florida to develop bids for our fixed-price spectrograph contract bids (including FLAMINGOS-2 at >$4M, and MIRADAS at >$11M). For the 1.6m demonstrator we estimate a total cost of $0.8M. Extrapolating from that estimate for the PolyOculus array, we estimate construction costs at $5M for the 5m-equivalent PolyOculus array, and an additional $0.5M for the spectrograph. With an additional $1.7M estimated for the 3-year operations phase (10% of the construction cost per year), we have a total cost of $8M (including the 1.6m demonstrator). All amounts are current-year dollars. We note that on longer operational timescales than this, the annual maintenance budget would increase over this amount by up to ~$0.3M/year in combined hardware and personnel, due to the need to replace individual COTS telescope units, as their MTBF will decrease over an extended lifetime.